# Cation-size mismatch as a predictive descriptor for structural distortion, configurational disorder, and valence-band splitting in II-IV-N$_2$ semiconductors

Malhar Kute,[1] Zihao Deng,[1] Sieun Chae,[1] and Emmanouil Kioupakis[1,*]

[1]*Department of Materials Science and Engineering, University of Michigan, Ann Arbor, 48109, United States*

*Correspondence to: kioup@umich.edu

*The II-IV-N$_2$ class of heterovalent ternary nitrides have gained significant interest as alternatives to the III-nitrides for electronic and optoelectronic applications. In this study, we apply first-principles calculations based on density functional theory to systematically investigate the effects of structural distortions due to cation size mismatch on the configurational disorder of the cation sublattice and the valence band structure in this class of materials. We find that larger size mismatch between the group-II and the group-IV cations results in stronger lattice distortions from the ideal hexagonal ratio, which in turn inhibits the propensity of these materials towards octet-rule violating cation disorder. We also demonstrate that the formation energy of a single cation antisite pair, which is fast and simple to calculate, is a strong indicator of a material's propensity towards disorder. Furthermore, the breaking of in-plane symmetry leads to a splitting of the top three valence bands at Γ, which is also directly related to the magnitude of structural distortions. Our work demonstrates that the structural and functional properties of the II-IV-N$_2$ materials can be finely tuned through controllable structural distortions that stem from the choice of cations.*

In recent years, the II-IV-N$_2$ class of materials have attracted significant interest as a structural analogue and potential alternative to the group-III nitride semiconductors. In their fully ordered phase, these materials adopt an orthorhombic structure (Pna2$_1$) derived from breaking the symmetry of the III-N wurtzite structure by substituting the group-III cations with alternating pairs of group-II and group-IV elements. However, the cation heterovalency adds an additional structural degree of freedom by enabling octet-rule-violating cation disorder, which can decrease the band gap by as much as 1-1.5 eV, indicating that the II-IV-N$_2$ materials could provide a broad range of band-gap tunability.[1–3] Many of the group-II and group-IV ions are much more sustainable and Earth-abundant than their group-III counterparts,[4] enabling their deployment in large-scale electronic and optoelectronic applications.

Many II-IV nitride materials have been synthesized since the 1960s, such as $BeSiN_2$,[5] $MgSiN_2$,[6] $MgGeN_2$,[6] $MgSnN_2$,[1,7,8] $ZnSiN_2$,[9] $ZnGeN_2$,[9–13] and $ZnSnN_2$.[3,14–21] The latter two have been recently grown as thin films using techniques such as MOCVD, MBE and sputtering, among others, at relatively low temperatures (e.g., <600°C for $ZnSnN_2$).[3,14–21] $MgSnN_2$ has been synthesized in both the rock salt[7] and wurtzite[1,8] phases. Despite significant advances in both theory and experiment, it is difficult to verify band-structure predictions experimentally due to disorder, off-stoichiometry,[2] unintentional dopants,[22,23] and the fact that several of these materials have not yet been synthesized. Given the numerous combinations of the group-II atoms beryllium, magnesium, zinc and cadmium, and the group-IV atoms silicon, germanium, and tin, there is a need to understand how the choice of cations affects the structural and electronic properties of heterovalent materials.

Monte Carlo simulations have shown that disorder becomes thermodynamically favorable above an effective temperature of ~2000°C for $ZnGeN_2$[24] and ~1000°C for $ZnSnN_2$,[2] much higher than the growth temperatures for either material, implying that the fully ordered crystal is the ground state, and that disorder results from kinetic trapping during non-equilibrium growth. Indeed, $ZnGeN_2$ and $ZnSnN_2$ are more disordered when grown at lower temperatures.[3,25] This contrasts with other II-IV-$V_2$ materials, which commonly exhibit a disorder transition temperature linked to the distortion of the hexagonal (sometimes called the tetragonal) plane, which in turn is related to the difference in the cation radii; distorted materials have a lower tendency to disorder than those that maintain their tetragonality.[26] Although this was established in the 1960s, this trend has not been explored for II-IV-N$_2$, perhaps because their disorder stems from kinetic trapping rather than thermodynamic stability. However, their disorder may still be linked to thermodynamic disorder temperatures, as ZnGeN$_2$ has been synthesized in both ordered and disordered forms, while ZnSnN$_2$ almost always appears disordered,[27] with only one growth report of the ordered phase,[14] correlating to their respective disorder temperatures.[2,24] Disorder in II-IV-N$_2$ has also been explored in terms of cation antisite pairs, which result in octet-rule violating disorder while maintaining stoichiometry and have been identified as a cause for band-gap reduction.[28,29]

Another key difference between III-N and their II-IV-N$_2$ counterparts is the splitting of the top three valence bands at the Γ point. For III-N, the top three bands are split by the crystalline anisotropy to a pair of nearly degenerate heavy and light hole bands (separated due to spin-orbit coupling by <10 meV)[30] that arise from the in-plane N $2p_x$ and $2p_y$ orbitals, and a third crystal-field band composed of N $2p_z$ orbitals.[23] In II-IV-N$_2$, the valence bands are further split into three distinct levels, due to the breaking of symmetry within the hexagonal plane by the two chemically distinct cations. One possible advantage of reducing the valence-band degeneracy for functional applications is that the holes at the valence band maximum can only scatter by phonon absorption into the same band if the energy splitting between the bands is larger than the thermal energy, k$_B$T (~25 meV at room temperature), thus suppressing inter-band hole scattering. Therefore, valence-band splitting greatly reduces the number of scattering channels for holes and has been previously shown to increase the hole mobility in strained GaN[31] and BAs.[32] Thus, II-IV-N$_2$ may have a higher phonon-limited hole mobility compared to III-N without the need for additional strain. Moreover, optical transitions between a singly degenerate valence band with N $2p$-orbital character and the conduction band with cation $s$-orbital character yield light emission polarized along the direction of the N $2p$ orbital.[33]

In this study we aim to explore how the heterovalency and cation-size mismatch in II-IV-$N_2$ affects the structural distortions, propensity towards disorder, and valence-band splitting. We find that a large difference in the group-II and group-IV cation radii results in significant in-plane lattice distortion. Both this cation size mismatch and the resulting distortions correlate to a higher antisite-pair formation energy, a lower propensity towards cation disorder, and increased splitting of the valence bands associated with the $p_x$ and $p_y$ orbitals. With a few exceptions, these materials behave in a consistent manner, indicating that simple elemental properties can be used to intuitively predict the material properties of the II-IV-$N_2$ class of materials relevant to optoelectronic and photovoltaic applications.

For this study, we performed first-principles calculations based on density functional theory (DFT) with the projector augmented wave method, as implemented in the Vienna ab-initio simulation package (VASP).[34,35] While some members of this material class are also stable in other crystal structures (e.g., rocksalt-type $MgSnN_2$[7]), we performed all calculations for the $Pna2_1$ structure, which is the stable structure for the majority of the investigated materials, to better establish consistent universal trends. We expect our qualitative conclusions to hold for the other stable structures as well. We used a plane-wave energy cutoff of 500 eV and a Γ-centered $4 \times 4 \times 4$ Brillouin-zone sampling grid for the 16-atom unit cells, scaled down to a $2 \times 2 \times 2$ mesh for the 128-atom supercells. We employed both the Perdew-Becke-Ernzerhof (PBE) functional[36] and the Heyd-Scuseria-Ernzerhof (HSE06) hybrid exchange-correlation functional[37] with a mixing parameter of α = 0.25. We fully relaxed the ionic positions and lattice parameters until the forces on the ions were less than $10^{-3}$ eV/Å for the ordered materials and $10^{-2}$ eV/Å for the disordered supercells, and we relaxed the ionic positions with fixed lattice parameters for the defective supercells until the forces on the ions were less than $10^{-2}$ eV/Å. Crystal structures shown in this report were created using the VESTA visualization software.[38]

The aliovalent nature of the cation sublattice of II-IV-$N_2$ inevitably leads to structural distortions from the perfect wurtzite structure. An intuitive metric to quantify the lattice distortions is the in-plane lattice deformation, defined as the stretching or compressing of the hexagonal-plane lattice constants compared to the ideal wurtzite structure. GaN and similar group-III nitrides are typically found in the wurtzite crystal structure, which when viewed in the orthorhombic basis, has an $a/b$ ratio of $\sqrt{3}/2$, where the lattice vectors are defined such that $b > a > c$. In this case, $a = \sqrt{3}a_w$ and $b = 2a_w$, where $a_w$ is the in-plane wurtzite lattice parameter; other conventions are occasionally used in the literature. Using the same convention for the orthorhombic II-IV-$N_2$ crystals, we observe that the adoption of two differently sized cations, as opposed to a singular cation as with III-N, results in a deviation from this ideal $a/b$ ratio, distorting the otherwise perfectly hexagonal pattern in the (0001) wurtzite basal plane. Interestingly, we find that this in-plane distortion is directly proportional to the difference in cation radii, here defined as $r_{II} - r_{IV}$, where $r_{II}$ and $r_{IV}$ are the Shannon ionic radii of the group-II and group-IV elements.[39] The relative in-plane distortion is calculated as:

$$\frac{\frac{a}{b} - \frac{\sqrt{3}}{2}}{\frac{\sqrt{3}}{2}} \times 100\%. \qquad (1)$$

The relationship between the cation radii difference and the relative in-plane distortion is shown in Figure 1. It is important to note that the data pass through the origin, indicating that a larger group-IV ion elongates the $a$ lattice parameter and compresses $b$, while a larger group-II ion instead decreases the $a/b$ ratio. The close, linear relationship between the in-plane distortion and the cation radius difference indicates that the structural distortion in II-IV-$N_2$ is a direct consequence of the size-mismatched cations, and both metrics can be used interchangeably as a measure to quantify lattice distortion. The full lattice parameters for each material can be found in Supplementary Table SI.

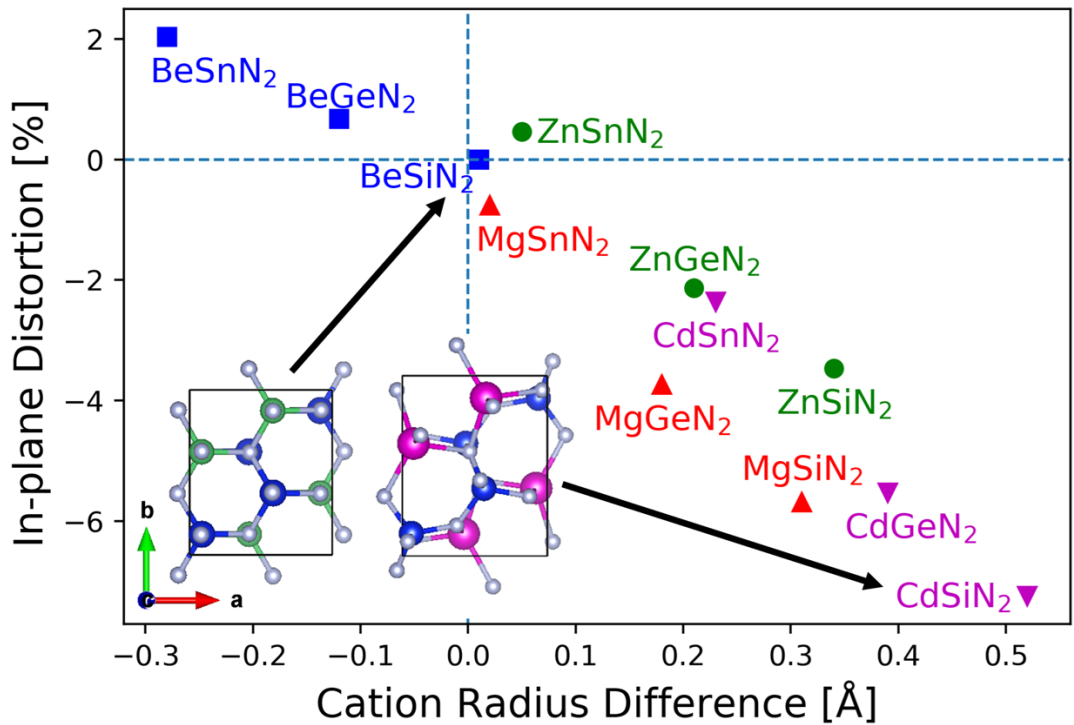

FIG. 1. Relative in-plane distortion, calculated according to Eq. 1, as a function of the ionic radius difference between the group-II and group-IV atoms in each material. The crystal structures shown indicate two extreme cases: the $BeSiN_2$ crystal (left) with almost no in-plane distortion (0.001%), and the $CdSiN_2$ crystal (right) with an in-plane distortion of -7.247%, scaled such that their $a$ lattice parameters are the same length to aid visual comparison.

Another key difference between II-IV-$N_2$ and their III-N counterparts is the possibility of exhibiting octet-rule violating configurational disorder in the cation sublattice due to their heterovalent nature. However, past works have shown that some materials within the II-IV-$N_2$ family (e.g., ZnSnN$_2$, MgSnN$_2$) have higher tendency to show disordering than others (e.g., ZnGeN$_2$). In this respect, it is crucial quantify each material's propensity towards a disordered cation sublattice. The first metric we propose is the slope of the formation energy for a series of supercells with increasing degree of cation disorder. We calculate the formation energy of 15 disordered 128-atom supercells, which are constructed by randomly swapping group-II and group-IV cations, resulting in different degrees of disorder. When the material is perfectly ordered, the cation sublattice consists only of the octet-rule-conserving (II)$_2$(IV)$_2$ N-centered tetrahedral motifs. However, if the material is disordered, (II)$_1$(IV)$_3$ and (II)$_3$(IV)$_1$ motifs start to populate the lattice, breaking the charge neutrality around the N atoms. We do not include (II)$_4$ and (IV)$_4$ motifs in this study, as they have a high formation enthalpy and are thus thermodynamically unfavorable to form, as has been shown, e.g., for ZnGeN$_2$[24] and ZnSnN$_2$[2]. Since stoichiometry is preserved in these supercells, the concentrations of (II)$_1$(IV)$_3$ and (II)$_3$(IV)$_1$ motifs are equal and can serve as the order parameter.

As expected, the formation energy of these supercells, calculated with the PBE functional and referenced to the ordered cell, increases linearly with the (II)$_1$(IV)$_3$-(II)$_3$(IV)$_1$ motif fraction, as shown for Zn-IV-N$_2$ in Figure 2. The slope of this curve varies for each material, with lower slopes indicating a lower energy penalty to reach a certain degree of disorder and leading to an increased likelihood for a material to exhibit disorder during experimental growth. These results match our expectations based on the order-disorder transition temperatures previously mentioned, which indicate a higher propensity for disorder for ZnSnN$_2$ than ZnGeN$_2$.[2,24] The deviations from linearity observed in Figure 2 and Supplementary Figure S1 are most likely due to interactions between antisite pairs. We also note that there is a slight tendency for more disordered supercells to have less deviation from the ideal wurtzite $a/b$ ratio. This phenomenon, shown in

Supplementary Figure S2, does not affect our conclusions about distortion in the ordered phase and these materials' tendency towards disorder.

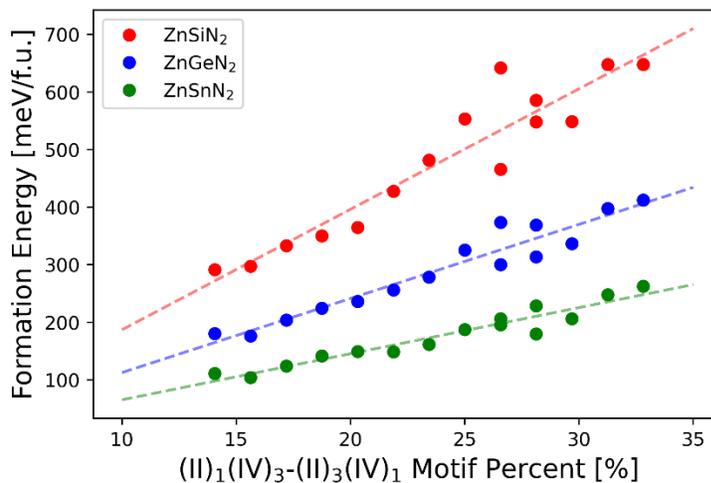

FIG. 2. The formation energy of disordered Zn-IV-$N_2$, referenced to the ordered bulk materials and calculated using the PBE functional, as a function of the $(II)_1(IV)_3$-$(II)_3(IV)_1$ motif fraction. The varying slopes indicate each material's propensity towards being disordered. Similar plots for the remaining materials in this study can be found in Supplementary Figure S1.

Even though the slope of the formation energy and $(II)_1(IV)_3$-$(II)_3(IV)_1$ motif percent gives a very good measure of disorder propensity, it requires calculations for several disordered supercells, which can be complicated and computationally expensive. Thus, we propose a simpler metric to quantify the tendency towards disordering for II-IV-$N_2$. Since the disorder supercell is constructed by progressively swapping cation pairs, the formation energy of a single nearest-neighbor cation swap, i.e., the antisite pair formation energy, can be viewed as a first-order approximation to the formation energy of disorder supercells. The antisite pair formation energy involves only one single supercell calculation, and it is therefore a more straightforward quantity to evaluate. We calculate the antisite pair formation energy using the formalism of point-defect calculations,[40] which for uncharged defects with no atoms removed or added simplifies to:

$$E^f[antisite\ pair] = E_{tot}[antisite\ pair] - E_{tot}[bulk], \quad (2)$$

where $E_{tot}[antisite\ pair]$ and $E_{tot}[bulk]$ are the total energy of the supercell containing the defect and the energy of the bulk material, respectively. We find that the choice of different nearest-neighbor antisite pairs affects the antisite formation energy by only a few percent (e.g., with an error bar of 4.00% for $BeGeN_2$ and 7.24% for $MgSiN_2$, as calculated within PBE for these two materials that span a broad range of structural distortions). Moreover, we find that second and third nearest-neighbor cation antisites have significantly higher formation energy than the first nearest neighbors (e.g., at least ~0.8 eV as calculated for $MgSiN_2$ within PBE, Fig. S3) and thus occur at much lower concentrations.

As expected, we find a strong linear correlation between the slope of the disordered supercell formation energy curves (Fig. 2) and the nearest-neighbor antisite formation energy, shown in Figure 3. This indicates that the antisite-pair formation energy can indeed be employed interchangeably with energy slopes, with lower values indicative of a stronger tendency towards cation disorder.

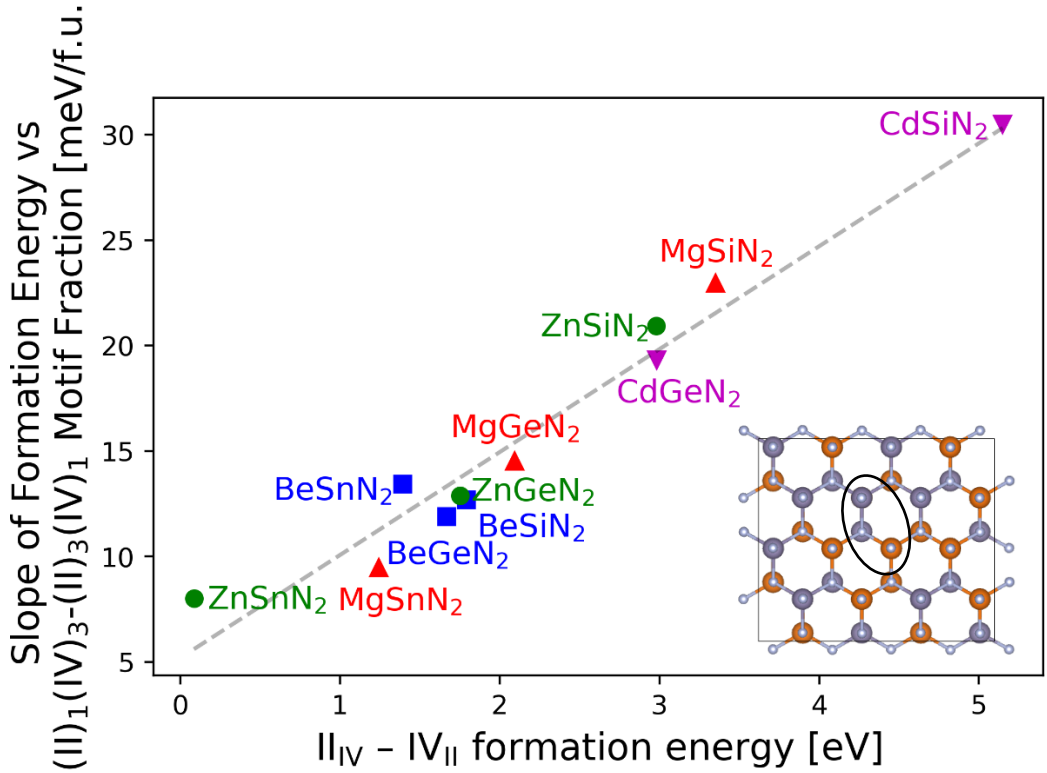

FIG. 3. The slope of the linear trend for the formation energy of 15 different disordered II-IV-$N_2$ supercells with varying $(II)_1(IV)_3$-$(II)_3(IV)_1$ motif fraction (e.g., Fig. 2) as a function of a single antisite pair formation energy, demonstrating a linear correlation between the two quantities. All energy values were calculated using the PBE functional. The inset crystal structure shows a supercell containing a cation antisite pair. The results show that the nearest-neighbor antisite pair formation energy is a good indicator for a material's propensity towards cation disorder.

In addition to establishing practical measures for lattice distortion and cation disordering, we also seek to establish whether compositional disorder directly correlates with inherent lattice distortion. Indeed, Figure 4 shows that the antisite formation energy is directly related to the absolute in-plane lattice distortion, and hence to the size mismatch of the two cations. Interestingly, each series of materials with a common group-II cation (e.g., Be-IV-$N_2$, Mg-IV-$N_2$) appears to follow slightly stronger linear trends, the slope of which correlates with the size of the group-II atom. The Be-IV-$N_2$ materials have a very small slope, roughly outlining a horizontal line, corresponding with the small ionic radius of beryllium (0.27 Å). Cadmium has a large ionic radius (0.78 Å), and the Cd-containing materials outline a steeper line than the trend for all 12 materials. Mg-IV-$N_2$ and Zn-IV-$N_2$ materials do not differ noticeably in their trends, as their group-II ionic radii are nearly identical (0.57 Å and 0.60 Å, respectively). The series formed by each group-IV cation (e.g., II-Si-$N_2$) do not display as strong trends as their group-II counterparts, most likely because the group-II ions are all larger than the group-IV ions in this study, apart from beryllium, which is approximately equal in size to Si (0.26 Å) and smaller than Ge and Sn (0.38 Å and 0.55 Å, respectively). Because of this exception, we note that Be-IV-$N_2$ does not strictly follow the positive linear trend of the other materials.

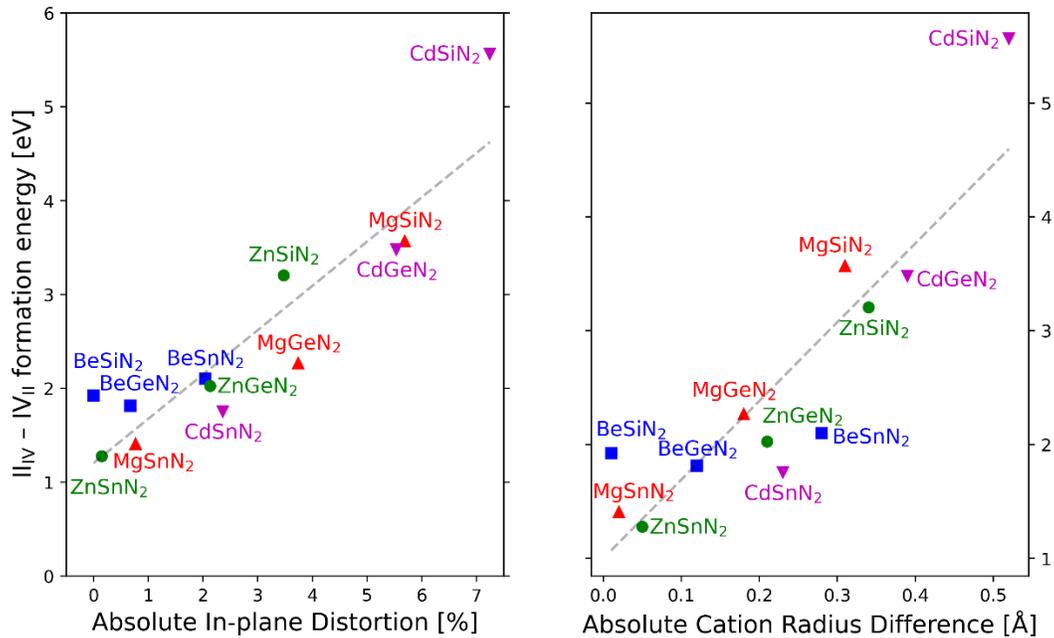

FIG. 4. The cation antisite pair formation energy plotted against the absolute value of in-plane distortion, calculated using Eq. 1 (left), and the absolute difference in the group-II and group-IV

ionic radii (right). The antisite pair formation energy was calculated using the HSE06 functional. Strong positive correlations exist between the lattice distortions and the disorder propensity in II-IV-N$_2$.

The cation size mismatch and associated in-plane distortions appear to also have a significant impact on the electronic properties. Specifically, the top three valence bands, two of which are nearly degenerate at the valence band maximum (Γ) for the group-III nitrides, split into three distinct energy levels for the II-IV nitrides, due to the symmetry breaking introduced by the in-plane distortion. All the materials in this study have direct band gaps at Γ, with the exception of Si-containing materials.[41–45] BeSiN$_2$ has its conduction band minimum between Y and Γ,[42,43] while MgSiN$_2$,[44] ZnSiN$_2$,[41] and CdSiN$_2$[43] have valence band maxima at T. The latter three do not exhibit the same valence-band splitting at their valence band maxima; the top two valence bands are degenerate at T. We consider the valence-band splitting at the Γ point, where the minimum direct band gap is located for all materials in this study other than BeSiN$_2$. Our results for valence-band splitting align reasonably well with those from other theoretical studies.[43,44] Hole effective masses have also been reported for most of these materials.[41,42,44,46] It is important to note that the valence bands do not split in the same manner for each material. For example, the top valence band at Γ for ZnSiN$_2$ and ZnGeN$_2$ primarily has $p_x$-orbital character, while for ZnSnN$_2$ this same band has $p_y$ character. Thus, we project the top three valence bands onto the $p_x$, $p_y$, and $p_z$ orbitals, which align with the $a$, $b$, and $c$ lattice parameters respectively. MgSiN$_2$ and CdSiN$_2$ exhibit mixed-orbital character in their top three bands and are thus excluded from Figure 5.

We find that the splitting between the $p_x$ and $p_y$ valence bands, which corresponds to the energy difference between the heavy and light hole bands in wurtzite group-III nitrides, correlates with the in-plane lattice distortion, as shown in Figure 5. The data pass approximately through the origin, indicating that an $a/b$ ratio greater than $\sqrt{3}/2$ results in a y-aligned valence band energy higher than the x-aligned valence band. Similarly, a smaller $a/b$ ratio results in $E_{p_x} > E_{p_y}$. The full results, including top valence-band polarization, can be seen in Supplementary Table SII. No notable trends were observed for the splitting associated with the $p_z$ valence bands. This variety in splitting behavior results in light polarization and inter-band scattering suppression that can be carefully engineered through the choice of cations, degree of disorder, or even applied strain.

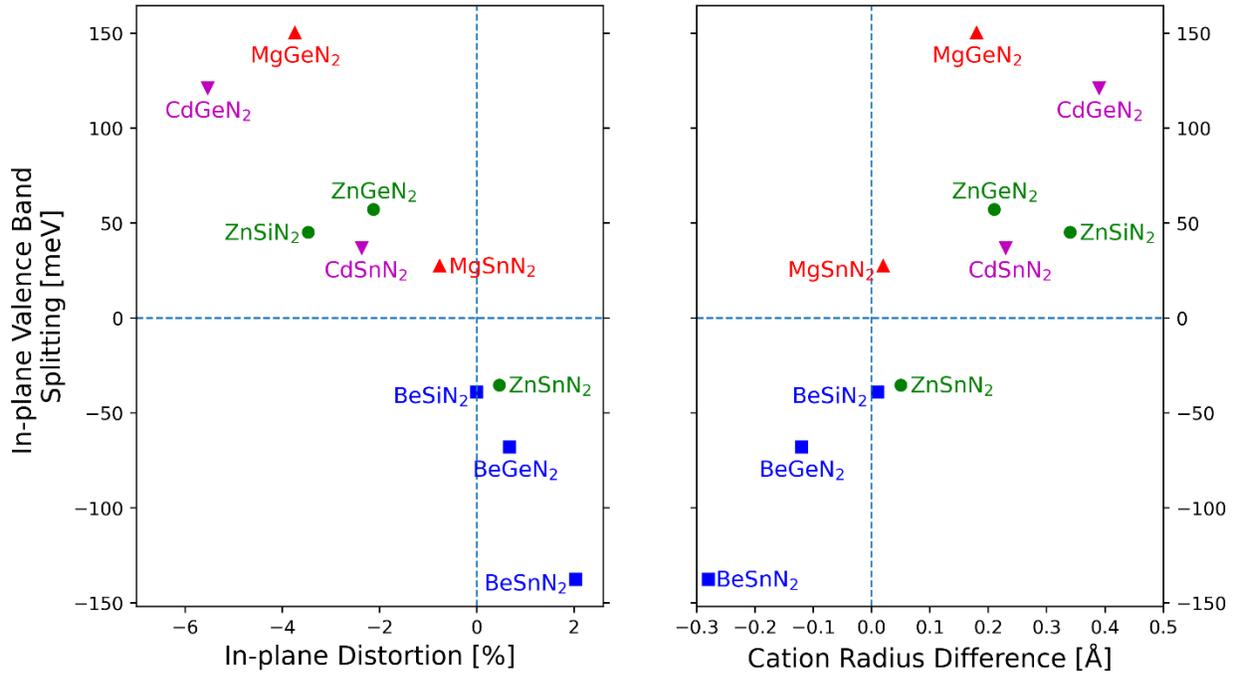

FIG. 5. Valence-band splitting between the top valence bands associated with the $p_x$ and $p_y$ orbitals at the $\Gamma$ point, $E_{p_x} - E_{p_y}$, plotted against the relative in-plane lattice distortion as defined in Eq. 1 (left) and the difference in cation radius, $r_{II} - r_{IV}$ (right). Increasing lattice distortions caused by differently sized cations result in an increase of the energy splitting of the $p_x$ and $p_y$ valence bands.

In conclusion, we establish a direct correlation between cation-size mismatch and lattice distortion, propensity towards cation disorder, and a splitting of the top three valence bands in II-IV-N$_2$ materials. Previous studies have shown how configurational disorder can be used to fine-tune the optical band gap for several of these materials,[1–3] and we show that this disorder can be controllably engineered by the choice of cations, as more similarly sized ions are more likely to form antisite defects. Furthermore, the difference in cation radii breaks the in-plane symmetry of the wurtzite structure and splits the valence bands, leading to polarized light emission and potentially suppressing inter-band scattering in comparison to the group-III nitrides. It is possible that the trends established in this study may be applicable to non-nitride II-IV-V$_2$ materials or other classes of heterovalent semiconductors. Our results show that structural distortions controlled by the choice of the two cations in II-IV-N$_2$ enable numerous advantages in electronic and optoelectronic applications compared to the conventional group-III nitrides.

See the supplementary material for the lattice parameters, formation energy of disordered supercells as a function of the $(II)_1(IV)_3$-$(II)_3(IV)_1$ motif percent, in-plane lattice distortion plotted against the $(II)_1(IV)_3$-$(II)_3(IV)_1$ motif percent, antisite formation energy as a function of cation swap distance, and the energy splittings and orbital characters of the top three valence bands for all investigated II-IV nitride materials.

**Acknowledgements**

This study was supported by the National Science Foundation through Grant No. DMR-1561008. It used resources of the National Energy Research Scientific Computing (NERSC) Center, a DOE Office of Science User Facility supported under Contract No. DE-AC02-05CH11231.

**DATA AVAILABILITY**

The data that support the findings of this study are available from the corresponding author upon reasonable request.

# Supplementary Information: Cation-size mismatch as a predictive descriptor for structural distortion, configurational disorder, and valence-band splitting in II-IV-N$_2$ semiconductors

Malhar Kute,[1] Zihao Deng,[1] Sieun Chae,[1] and Emmanouil Kioupakis[1,*]

[1]*Department of Materials Science and Engineering, University of Michigan, Ann Arbor, 48109, United States*

*Correspondence to: kioup@umich.edu

TABLE SI. Lattice parameters for II-IV-N$_2$ materials in the orthorhombic Pna2$_1$ crystal structure, calculated using the HSE06 functional.

| Material | a | b | c |
|---|---|---|---|
| BeSiN$_2$ | 4.9606 | 5.7281 | 4.6591 |
| BeGeN$_2$ | 5.1517 | 5.9090 | 4.8368 |
| BeSnN$_2$ | 5.4920 | 6.2151 | 5.0808 |
| MgSiN$_2$ | 5.2616 | 6.4422 | 4.9767 |
| MgGeN$_2$ | 5.5083 | 6.6077 | 5.1779 |
| MgSnN$_2$ | 5.9021 | 6.8678 | 5.4584 |
| ZnSiN$_2$ | 5.2393 | 6.2674 | 5.0215 |
| ZnGeN$_2$ | 5.4719 | 6.4559 | 5.2025 |
| ZnSnN$_2$ | 5.8529 | 6.7485 | 5.4680 |
| CdSiN$_2$ | 5.3851 | 6.7040 | 5.2471 |
| CdGeN$_2$ | 5.6418 | 6.8962 | 5.4396 |
| CdSnN$_2$ | 6.0646 | 7.1723 | 5.7256 |

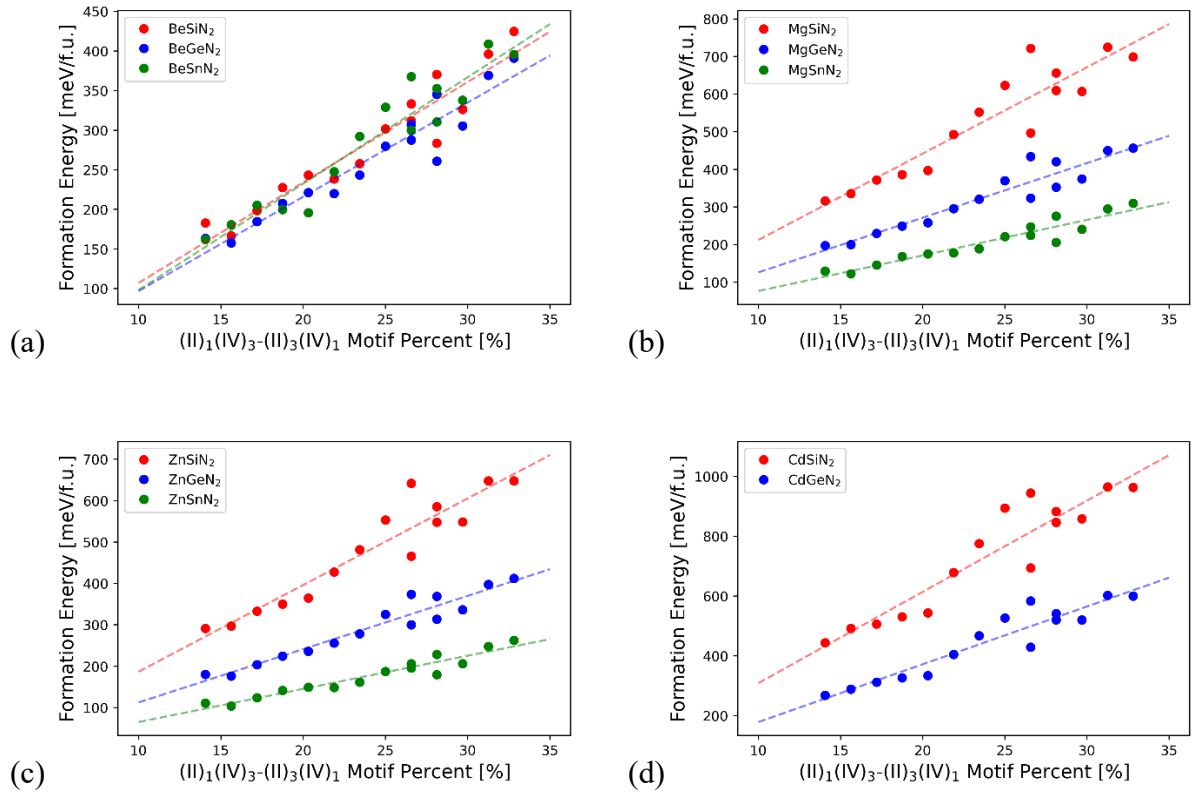

FIG. S1. Formation energy of disordered 128-atom supercells plotted against the $(II)_1(IV)_3$-$(II)_3(IV)_1$ motif percent of each cell. The slope of each curve indicates the propensity of a material to form a disordered structure and is directly related to the antisite formation energy. Deviation from the linear trend is likely due to interactions between charged octet rule-violating motifs.

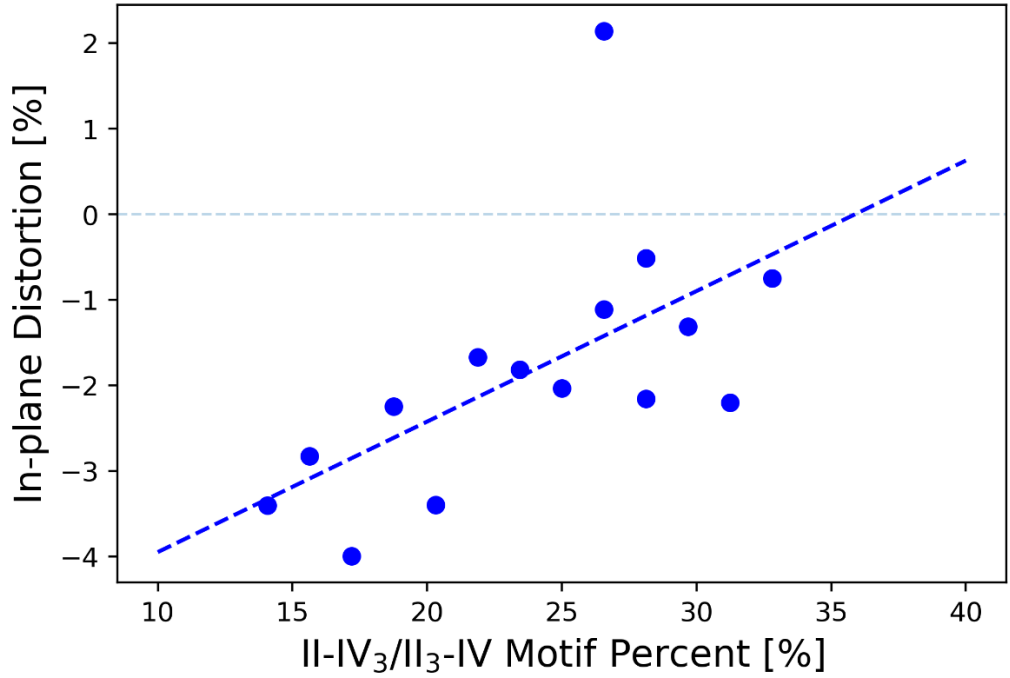

FIG. S2. In-plane lattice distortion, calculated according to Eq. 1, of the 15 disordered supercells for MgSiN$_2$, plotted against the $(II)_1(IV)_3$-$(II)_3(IV)_1$ motif fraction of the supercells. As the supercells become more disordered, the distortion tends towards the ideal wurtzite symmetry. The deviations from linearity are likely due to the relatively small supercell size.

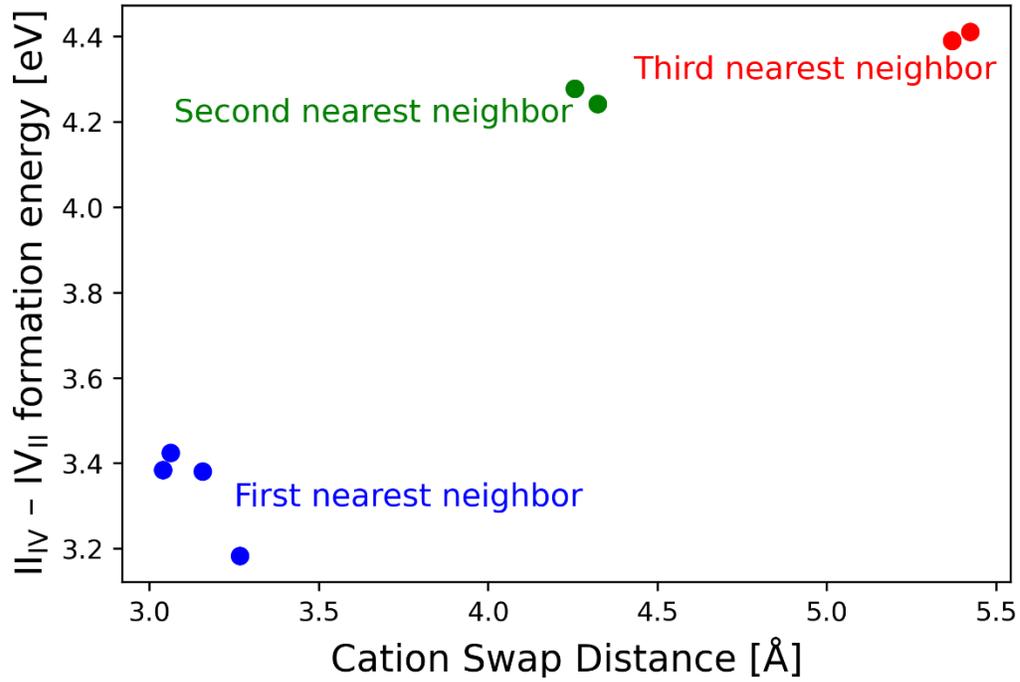

FIG. S3. The cation antisite formation energy for MgSiN$_2$ increases with increasing distance between the two cation sites. This indicates that nearest-neighbor swaps are energetically more favorable than antisites farther apart, and that (II)(IV)$_3$-(II)$_3$(IV) motifs have lower energy in close proximity to each other. These calculations were performed with the PBE functional.

TABLE SII. Valence-band splitting at Γ for all materials in this study. Here, $\Delta E_{12}$ and $\Delta E_{23}$ refer to the different in energy of the top two valence bands and the second and third highest valence bands, respectively, without consideration for their orbital character. The orbital projection is also shown here for the top three valence bands in order of descending energy. MgSiN$_2$ and CdSiN$_2$ do not have clearly defined orbital character for their top valence bands.

| Material | $\Delta E_{12}$ [meV] | $\Delta E_{23}$ [meV] | Orbital Projection [1, 2, 3] |
|---|---|---|---|
| BeSiN$_2$ | 32 | 39 | $p_z, p_y, p_x$ |
| BeGeN$_2$ | 68 | 15 | $p_y, p_x, p_z$ |
| BeSnN$_2$ | 127 | 11 | $p_y, p_z, p_x$ |
| MgSiN$_2$ | 63 | 92 | --, $p_z, p_y$ |
| MgGeN$_2$ | 96 | 150 | $p_z, p_x, p_y$ |
| MgSnN$_2$ | 118 | 27 | $p_z, p_x, p_y$ |
| ZnSiN$_2$ | 45 | 25 | $p_x, p_y, p_z$ |
| ZnGeN$_2$ | 57 | 15 | $p_x, p_y, p_z$ |
| ZnSnN$_2$ | 35 | 14 | $p_y, p_x, p_z$ |
| CdSiN$_2$ | 56 | 240 | --, $p_y, p_x$ |
| CdGeN$_2$ | 106 | 15 | $p_x, p_z, p_y$ |
| CdSnN$_2$ | 37 | 19 | $p_x, p_y, p_z$ |